\documentclass[journal]{IEEEtran}
\usepackage{cite}

\ifCLASSINFOpdf
   \usepackage[pdftex]{graphicx}
\else
   \usepackage[dvips]{graphicx}
\fi
\usepackage{amsmath}

\usepackage{amsfonts}
\usepackage{amssymb}
\usepackage{tabularx}

\usepackage{algorithmic}
\usepackage{physics}

\usepackage{array}

\newcounter{MYtempeqncnt}

\begin{document}
%
\title{Joint Access Point Selection and Interference 
\\ Cancellation for Cell-Free Massive MIMO}
%
%
%

\author{Indu L.~Shakya,
        and Falah H.~Ali,~\IEEEmembership{Senior Member,~IEEE,}
\thanks{Indu L. Shakya was with the University of Sussex, Brighton, BN1 9QT UK.
He is now a technology leader and consultant for wireless and satellite
communications industries (ishakya@gmail.com), Falah H. Ali is a Professor of Communications Engineering, at the School of Engineering and Informatics,
University of Sussex, Brighton, BN1 9QT UK. (f.h.ali@sussex.ac.uk).}}

%
%

\markboth{IEEE Journal,~Vol.~xx, No.~xx, yy~202X}%
{Shell \MakeLowercase{\textit{et al.}}: Bare Demo of IEEEtran.cls for IEEE Journals}
%



\maketitle

\begin{abstract}
Cell-Free Massive MIMO is a highly promising approach to enhance network capacity by moving a large number of distributed access points (AP) closer to mobile users while utilizing
simple matched filtering and conjugate beamforming. Recent work using minimum mean-squared-error (MMSE) receiver that suppress multi-user interference (MUI) shows significant capacity increase, but at the cost of high computational complexity and residual MUI enhancement. We propose a significantly lower complexity adaptive approach where central processing unit (CPU) removes MUI without amplifying the residual interference. It does so dynamically by using available knowledge of channel estimates to perform joint process of combining selected strongest AP signals for each user and subtracting the sum of interference estimates from other users at the same time. We provide signal-to-interference plus noise-ratio (SINR) and complexity analyses backed by numerical results to show the superiority of this approach compared with the state-of-the-art techniques.
\end{abstract}


%
\IEEEpeerreviewmaketitle

\section{Introduction}
%
%
%
%
\IEEEPARstart{R}{ecently,} cell free (CF) massive MIMO has received widespread attention as being one of the most promising
approaches to enhance the user experience of mobile users for beyond 5G technologies by deploying a large number of
distributed antennas or APs closer to mobile users \cite{IEEEngo:cfmr}-\cite{IEEEicc:JPap}.  
User centric (UC) approach is detailed in \cite{IEEEbuzzi:cfuc} to select only a subset of APs  with strongest channel gains to reduce computations and backhaul signalling compared with full APs selection in \cite{IEEEngo:cfmr}. The AP selection scheme in \cite{IEEEcf:chg} uses only large-scale fading to associate each single antenna user to an AP, where each AP is equipped with more antennas than total users. Both \cite{IEEEbuzzi:cfuc} and \cite{IEEEcf:chg} utilize traditional interference ignorant matched filter (MF) detection method and hence achieve very low spectral efficiency from large array of distributed antennas. 

On the other hand,  MMSE receivers with interference suppression capabilities at the CPU is investigated in \cite{IEEEbjo:cfmmse} to show substantial capacity gains while ignoring computational load aspects. They require inversion of large matrices, and hence significantly higher computations to afford the gains in capacity. Two practical aspects and crucial insights for CF massive MIMO we highlight here are: a) most of the useful energies from each user's transmission are distributed around nearby APs but these are subject to change all the time, and, b) this being the case there is no big advantage from wasting computational resources to suppress interference from the APs whose energies are too low. Therefore, a more practical and adaptive approach is needed to address the interference issue without substantially increasing computational complexity. 

 
We address these aspects with a new design referred to as Joint AP Selection and Interference Cancellation (JAPSIC) that specifically, a)  combines signals only from the APs with strongest gains dynamically to generate better initial data estimates and b) within the same process, cancels sum of all interfering user’ estimates obtained in parallel from the raw data to refine desired users’ data estimates. The whole process  adds only a modest complexity over the benchmark MF method \cite{IEEEngo:cfmr} even after repeating over many iterations. Note that this approach should not be confused with the well-known successive/serial interference cancellation (SIC) — though can be seen as an adaptive version of the parallel IC (PIC) method used for CDMA and massive MIMO \cite{IETsha:ccdma},\cite{IEEEshen:mblast}.
Furthermore, it has an attractive feature of not amplifying residual interference and noise which the MMSE based schemes \cite{IEEEbjo:cfmmse} suffer from.
We derive SINR of JAPSIC and highlight how it offers an advantage over the MMSE under practical channel estimation error conditions. Numerical results are provided to show the gains against the alternatives \cite{IEEEngo:cfmr}, \cite{IEEEbuzzi:cfuc}, \cite{IEEEbjo:cfmmse} in terms of sum spectral efficiencies (SE), computational efforts, backhaul overheads to justify its attractiveness for implementation.

Notations: Bold faces lowercase letters $\textbf{x}$ denote column vectors; boldface uppercase letters $\textbf{X}$ denote matrices. The
superscripts $\{.\}^{T}$ and $\{.\}^{H}$ denote transpose and conjugate transpose, respectively; $\textbf{0}_{N}$ denotes a row vector of size $N$
consisting of all zeros, $\textbf{I}_{M}$ denotes an identity matrix of size ${M} \times {M}$ . The operator $\mathbb E\{x\}$ denotes expectation with respect to
$\{x\}$; and size$\{.\}$ denotes the cardinality of the input data.
\section{System Model}
We consider an uplink of an urban CF wireless environment with $K$ single antenna mobile users and $M$ distributed single antenna
APs that are connected via backhaul links to the CPU where all channel estimation and
decoding of users’ data is performed. 
We use a centralized setup similar to \cite{IEEEbjo:cfmmse} consisting of a)
training phase of pilot transmission from the mobile users to the APs to allow channel estimation at the CPU and b)
uplink data transmission phase from all mobile users to all APs generating complex raw data statistics that are sent
to the CPU for final decoding.

\subsection{Propagation Model}
The propagation model used here is based on 3GPP Urban Microcell model that captures the essence
of typical dense urban environment better than the three-slope path loss model \cite{IEEEbjo:cfmmse}. For a typical carrier
frequency of 2 GHz, this gives coefficients $\beta_{k,m}$, in dB, capturing large scale fading and shadowing effects as follows:
$\beta_{k,m}=-30.5 -36.7 \log 10\{\frac{d_{k}m}{1 m}\}+F_{k,m}$, where $d_{k,m}$ is the distance between user $k$ and AP $m$ and $F_{k,m}$ is the
shadow fading loss with distribution $\mathcal{N}\left(0,4^{2}\right)$. Mean values for shadowing correlation between APs and users are
assumed $0$ if they are spaced $> 50$ m apart and $4^{2}2^{-\delta_{k,i}/9m}$ otherwise, where $\delta_{k,i}$ is the distance between users $k$ and $i$.
Short-term fading terms $h_{k,m}$ are assumed flat across the coherence bandwidth and static during
each channel realization with coherence time of $\tau_{c}$ and follow Rayleigh distribution $\mathcal {CN}(0,1)$. The complex channel $g_{k,m}$ between user $k$ and
AP $m$ incorporating large scale fading, shadowing and short term fading can then be written as: 
$g_{k,m}=\beta_{k,m}^{1/2}h_{k,m}, k=1,2,.,K; m=1,2,.,M$.
\subsection{Uplink Training and Channel Estimation Model}
In this phase, all $K$ users simultaneously transmit pilot sequences of length $\tau_{p}$ to APs, that are forwarded to the CPU for estimating all the channels. We assume $\tau_{p}$ orthogonal pilot sequences
are used and high loading scenario of $\tau_{p}<<K$ so multiple users in a set $\mathcal{P}_{k}\subset \{1,2,.,K\},$ share the same sequence with $\rho=$ size$\{\mathcal{P}_{k}\}$ and MMSE
channel estimation is used. This leads to pilot contamination due to mutual interference of users degrading the channel estimation performance. A channel estimate obtained at the CPU for user $k$ at AP $m$, $g_{k,m}$ can be modelled as: $\hat{g}_{k,m}=g_{k,m}+c_{k,m}$, where $c_{k,m}$ is a channel estimation error that is uncorrelated with the channel gains, with distribution $\mathcal{CN}(0,\sigma^{2}_{c})$.
\subsection{Uplink Data Transmission}
In this phase lasting $\tau_{c}-\tau_{p}$ symbols, all $K$ users transmit their data $s_{k},k=1,2,.,K$ over their respective channels $g_{k,m}$ to give each AP
$m=1,2,.,M$ their received signals $y_{m}=\sum_{k=1}^{K}\sqrt{p_{k}}g_{k,m}s_{k}+z_{m}$. All APs then forward the signals to the CPU for final decoding. 
We assume complex Gaussian data symbols $s_{k}$ with max power $p_{k}$ for each user. The raw data vector collected at
the CPU from all APs at every symbol period can be shown in compact form as:
\begin{equation}
\label{eqn_y}
\mathbf{y} = \sum\limits_{k=1}^{K}  \mathbf{P}^{1/2} \mathbf{g}_{k} s_{k} + \mathbf{\mathbf{z}},
\end{equation}
where $\textbf{y} = [y_{1},y_{2}, . . , y_{M}]^{T} $, $\textbf{P} =$ diag$ [p_{1},p_{2}, . , p_{K}]$, $\textbf{g}_{k} =[g_{k,1}, g_{k,2}, . , g_{k,M}]^{T}$,
and ${\textbf{z}} = [z_{1},z_{2}, . , z_{M}]^{T}$ with $z_{m}$ representing additive thermal noise at each AP with $\mathcal{C}\mathcal{N}(0,\sigma^{2}_{z})$.

\section{Proposed JAPSIC Receiver}
The process for obtaining $k_{th}$ user’s data estimate $\hat{s}_{k}(n), k=1,2,.,K$ at the $n_{th}$
symbol period, $n = 1, 2, . , N$, involves taking the vector
$\textbf{y}(n)$ and multiplying it with a combining vector $\textbf{w}_k(n)$ as
follows:
\begin{equation}
\label{eqn_sk}
\hat{s}_{k}(n) = \mathbf{w}_{k}(n) \mathbf{y}(n).
\end{equation}
The data estimation process at the CPU involves specification of $\textbf{w}_{k}(n)$ depending on the receiver methods used, including
manipulations of intermediate soft estimates to arrive as close as possible to the original data $\hat{s}_{k}(n) \rightarrow s_{k}(n), \forall k, \forall n $.
\subsection{Existing CF Massive MIMO Schemes}
With the CF scheme using MF decoding \cite{IEEEngo:cfmr}, $m_{th}$ an AP sends soft data estimate that it obtains by multiplying the raw data with the conjugate of local channel
estimates i.e. $w_{k,m} = {g}^{*}_{k,m}$ while ignoring the presence of other users'
MUI contributions: $y_{k,m} = w_{k,m}y_{m}$. The CPU receives $M$ such
estimates to generate final data estimate $\hat{s}_{k} =\sum \limits_{m=1}^{M}y_{k,m}$. The UC approach \cite{IEEEbuzzi:cfuc} is obtained by processing subset of $M_{u} $ APs instead of all APs. The MMSE schemes suppress MUI and noise to  give better performance
than MF \cite{IEEEbjo:cfmmse}. This involves inverting channel matrix $\textbf{G}$,
estimation error $\textbf{C}$ and noise estimation $\sigma^{2}\textbf{I}_{M}$ matrices of sizes $K \times M$,
$K \times M$ and $M\times M$ to minimize the mean squared
error of data $\mathbb {E} \{\abs{ \hat{s}_{k}(n)- s_{k}(n)}\}^{2}$. The combining vector \cite{IEEEbjo:cfmmse},
dropping $(n)$ notation for simplicity here, can be written as:
\begin{equation}
\label{eqn_wk}
\mathbf{w}_{k} =p_{k}\Bigg(\sum \limits_{i=1}^{K} p_{i}\Big(\mathbf{\hat{g}}_{i}\mathbf{\hat{g}}_{i}^{H} +\mathbf{C}_{i}\Big)+\sigma^{2}\mathbf{I}_{M}\Bigg)^{-1}\mathbf{\hat{g}}^{H}_{k}.
\end{equation}
MMSE-SIC enhances upon MMSE by successively decoding and cancelling strongest users before decoding weaker users. 
\vspace{-3mm}
\subsection{Proposed JAPSIC Algorithms}
With JAPSIC, the CPU collects only selected APs' data from $\textbf{y}$ at each stage $l,l=0,2,.,L$ and exploits already available knowledge of all
users’ channel estimates $\hat{\textbf{g}}_k, k=1,2,.,K$, to generate cleaner data $\textbf{y}^{l}_{k}$ to refine desired user’s data estimate $\hat{s}^{l}_{k}$
in parallel. This entails taking the estimate from the previous stage $\hat{s}^{l-1}_{k}$ and subtracting sum of all interfering users’ estimates $\Psi^{l}_{k}$
from $\textbf{y}$ over $L$ stages/iterations while involving minimal divisions/multiplications. This gives JAPSIC a big advantage over MMSE in
computational efforts so they can be implementable even in highly mobile users' channel environments \cite{Wileyses:lte}. 
Two variants are detailed here:

\subsubsection{JAPSIC $\theta$ Algorithm} 
This variant utilizes a threshold value $\theta$ that is compared against each user’s estimated channel power at each AP $\abs{ \hat{g}_{k,m}}^{2}$, to use as a measure to qualify an
AP’s raw data for processing and cancellation in subsequent stages. The algorithm steps are shown in Table I.
\begin{table}[!t]
\renewcommand{\arraystretch}{1.3}
\caption{JAPSIC $\theta$ Algorithm For Data Estimation}
\label{table_JAPSIC_theta}
\centering
\begin{tabular}{l}
\hline
1) Set a channel-power threshold = $\theta, \forall k,\forall m; k=1, 2, . , K; m = $\\
$1,2, . , M $\\
2) For each channel coherence block $n=1:N$,\\
3) For each user, $k=1:K,$\\
4) Calculate indices vector ${\bf{\iota}}_{k}(n)$ by evaluating the threshold:\\
${\bf {\iota}}_{k}(n)=\abs{ {\bf{\hat{g}}}_{k}(n) }^{2} \geq \theta, \iota_{k}(n) \in \{1,2,.,M\}, $ \\
$\mu_{k}(n)= $ size$\{{\bf{\iota}}_{k}(n)\}\leq M, \forall k,$ \\
5) For each stage of detection, $l= 0, 1,2,.,L$,\\
\hspace{0.3cm} If $l=0; {\bf{w_{\iota}}}_{k}(n)={\bf{\hat{g_{\iota}}}}^{H}_{k}(n), \hat{s}^{0}_{k}(n)={\bf{w_{\iota}}}_{k}(n){\bf{y_{\iota}}}_{k}(n)$; else,\\
\hline
\hspace{0.3cm}	a) Obtain combining vector and raw data for the user by selecting\\
\hspace{0.3cm}	the subset with indices $\iota_{k}(n)$ from ${\bf{w}}_{k}(n)$: ${\bf{w}}_{\iota_{k}}(n)={\hat{\bf{g}}^{H}_{\iota_{k}}}(n)$.\\
\hline
\hspace{0.3cm}	b) Obtain the JAPSIC cancellation vector, $\Psi_{k}(n)$, by summing \\
\hspace{0.3cm}	all interfering users’ signals: $\Psi_{k}(n)=\sum \limits_{i, i\neq k}^{K}\hat{s}^{l-1}_{\iota}(n) \hat{\bf{g}}_{i \iota_{k}}(n)$.\\
\hspace{0.3cm}	c) Update data statistics for user $k$;
\hspace{0.3cm}	${\bf{y}}^{l}_{\iota_{k}}(n)={\bf{y_{\iota}}}_{k}(n)-\Psi_{k}(n)$.\\
\hline
\hspace{0.3cm}	d) Obtain a soft data estimate for the $k_{th}$ user $\hat{s}^{l}_{k}(n)$, using\\
\hspace{0.3cm}	signal statistics from all APs, $\hat{s}^{l}_{k}(n)={\bf{w}}_{\iota_{k}}(n){\bf{y}}^{l}_{\iota_{k}}(n)$.\\
\hline
6) Calculate SINR using (\ref{eqn_dbl_y}). End $k$, and end $n$.\\
7) Calculate mean number of APs selected $\mathcal{M}=\frac{1}{KN} \sum \limits_{k=1}^{K}\sum \limits_{n=1}^{N}\mu_{k}(n)$.\\
\hline
\end{tabular}
\end{table}
\subsubsection{JAPSIC $M_{u}$ Algorithm}
Here a fixed number of $M_{u} \leq M$ APs with strongest channel power is selected using the users’ channel estimates vectors from Phase a). See Table II.

\begin{table}[!t]
\renewcommand{\arraystretch}{1.3}
\caption{JAPSIC $M_{u}$ Algorithm For Data Estimation}
\label{table_JAPSIC_theta}
\centering
\begin{tabular}{l}
\hline
1) Set the number of APs to select $=M_{u}, \forall k, k=1,2,.,K, M_{u}\leq M$.\\
2) For each channel coherence block $n=1,2,.,N$,\\
3) For each user, $k=1,2,.,K$,\\
\hline
4) Initialize a vector of indices to be assigned to $M_{u}$ selected APs from \\
all $M$ APs,  ${\bf \iota}_{k}(n)={\bf 0}_{M_{u}}; M_{u} =$ size$\{{\bf{\iota}}_{k}(n) \} \leq M, \forall k$. \\
\hline
5) Derive the indices vector ${\bf{\iota}}_{k}(n)$, by sorting all APs' channel powers\\
$\abs{{\bf{g}}_{k}(n)}^{2}$ in descending order and picking only the first $M_{u}$ APs’ \\
indices.\\
\hline
6) Use the steps 5 and 6 as in Table I.\\
\hline
\end{tabular}
\vspace{-4mm}
\end{table}
Comparing the AP selection approaches of the two, JAPSIC $\theta$ is less complex in that it does not require sorting of APs, but it requires knowledge of their channel power ranges. Note the algorithms in Table I and II are highly amenable for optimization where an objective can be set e.g. to maximize the SINR in (\ref{eqn_dbl_y}) by gradually increasing $L$ while allowing $\theta$/$M_{u}$ to adapt until a desired or the peak SINR is found. 
\vspace{-1mm}
\section{SINR and Computational Complexity Analyses}
We start with derivation of SINR for the JAPSIC $\theta$ process at stage $0$ which consists of matched filtering of $\mu_{k}\leq M, k=1,2,.,K$ APs. The SINR for a user $k, \Gamma^{0}_{k},$ can be given as:
\begin{equation}
\label{eqn_stage0}
\Gamma^{0}_{k} \\
=\frac{p_{k}\abs{{\sum\limits_{m\in\bf{\iota}_{k}} }\hat{g}^{*}_{k,m}g_{k,m}}^{2}}   {\sum \limits_{i=1, i\neq k}^{K}p_{i}\abs{{\sum\limits_{m\in\bf{\iota}_{k}} } \hat{g}^{*}_{k,m}g_{i,m}}^{2}+\sigma^{2}_{z} {\sum\limits_{m\in\bf{\iota}_{k}} } \abs{\hat{g}_{k,m}}^{2}}.
\end{equation}

\begin{figure*}[!t]
\normalsize
\setcounter{MYtempeqncnt}{\value{equation}}
\setcounter{equation}{4}
\begin{equation}
\label{eqn_dbl_y}
\Gamma^{l}_{k} =\frac{p_{k}\abs{{\sum\limits_{m\in\bf{\iota}_{k}} }g^{*}_{k,m}g_{k,m}}^{2}}   {\underbrace{\sum \limits_{i=1, i\neq k}^{K}p_{i}\abs{{\sum\limits_{m\in\bf{\iota}_{k}} } s_{i}g^{*}_{k,m}g_{i,m}-\hat{s}^{l-1}_{i}g^{*}_{k,m}g_{i,m}}^{2}}_{\mathrm{IC}}+\underbrace{\sum \limits_{k=1, i=1}^{K}p_{k}\abs{{\sum\limits_{m\in\bf{\iota}_{k}} } \{\hat{g}_{k,m}-g_{k,m}\}\{\hat{g}_{i,m}-g_{i,m}\}}^{2}}_{\mathrm{RMUI}}+\sigma^{2}_{z} {\sum\limits_{m\in\bf{\iota}_{k}} } \abs{\hat{g}_{k,m}}^{2}}.
\end{equation}
\hrulefill
\vspace*{-5mm}
\end{figure*}
Using the raw data estimates from all users at stage $0$, $\hat{s}_{i}, \{i=1,2,.,K\}$, each subsequent stage $l$ of the JAPSIC $\theta$ algorithm refines the data estimates for each user by cancelling sum of all interefering users' estimates in parallel which is shown in equation (\ref{eqn_dbl_y}) on the next page. Here the numerator is formed by collecting energies from the desired user from the subset of $\mu_{k}(n) \leq M$ APs. The denominator is formed of interference cancellation (IC) output, residual MUI (RMUI) consisting of the sum of channel estimation error correlations and thermal noise from $\mu_{k}(n)$ APs. After sufficiently large iterations of cancelling the reconstructed MUI signals to refining desired user data, at final $L_{th}$ stage, the soft estimates consist of desired data affected by only RMUI and the total noise component. Note that linear interference cancellation methods that do not use hard decision in each stage such as one utilized here, refine data estimates without causing error propagation. Hence close to MUI free decoding can be achieved with e.g. $L=10$ stages \cite{IEEEshen:mblast}, \cite{IETsha:ccdma}. 
Finally, the achievable sum SE $\Upsilon_{sum}$, is obtained by
summing SEs of all $K$ users using their expected SINR values over all channel realizations:
\begin{equation}
\label{eqn_se}
\Upsilon_{sum} =\sum\limits_{k=1}^{K}\Big(1-\frac{\tau_{p}}{\tau_{c}}\Big) \mathbb{E}\Big\{ \log_{2} (1+\Gamma_{k})\Big\}.
\end{equation}

Next, we analyse and compare SINR of JAPSIC against MMSE to assess its robustness against possible pilot contamination (we use JAPSIC $M_{u}$ for the ease of presentation). We add an estimation error correlation matrix $\mathbf{C}$ into the channel matrix $\mathbf{\hat{G}}$ and obtain SINR expressions. For MMSE receivers this can be obtained as follows for a $k^{th}$ user \cite{IEEEngo:mmse}:
\begin{eqnarray}
\label{sinr_mmse_a}
\Gamma^{\mathrm{MMSE}}_{k} =\frac{1}{1-{\Bigg\{{\mathbf{\hat{G}}}^{H}\Big({\sigma^{2}_{z}{\mathbf{I}_{M}}+{\mathbf{\hat{G}}}{\mathbf{\hat{G}}}^{H}+\mathbf{C}}\Big)^{-1}\mathbf{\hat{G}}\Bigg\}_{k,k}}}.  
\end{eqnarray}
Equivalently, an SINR for JAPSIC $M_{u}$ assuming large enough IC stages $L$ clearing the MUI component, can be given as:
\begin{eqnarray}
\label{sinr_japsic_a}
\Gamma^{\mathrm{JAPSIC}}_{k}=\lim_{{\abs{s_{i}-\hat{s}^{L}_{i}}^{2} \rightarrow 0}, \forall i\neq k} \frac{\Big\{{\mathbf{\hat{G}}_{M_u}^{H}{\mathbf{\hat{G}}_{M_u}}}\Big\}_{k,k}}{\Big\{{{\sigma^{2}_{z}{\mathbf{I}}_{M_{u}}+\mathbf{C}_{M_{u}}}}\Big\}_{k,k}},
\end{eqnarray}

where $\mathbf{\hat{G}}_{M_u}$ and $\mathbf{C}_{M_{u}}$ are obtained by picking $M_{u}$ rows from $\mathbf{\hat{G}}$ and $\mathbf{C}$, respectively. While comparing  (\ref{sinr_mmse_a}) and (\ref{sinr_japsic_a}), we can not make definite conclusions about their relative superiorities \textrm {--} we anticipate that at higher SNRs, the SINR loss due to channel estimation error enhancement of MMSE \cite{IETsha:ccdma}, \cite{IEEEshen:mblast} will be more visible. Note that MMSE-SIC does not offer much gain over MMSE for higher $M$ as SIC is less effective in removing RMUI due to channel hardening  \cite{IEEEbjo:cfmmse}.

We also analyze and compare the computational efforts of JAPSIC with the others in terms of complex multiplication/division operations while ignoring additions and subtraction terms in Table III. All schemes include matched filtering at the initial stage. The JAPSIC $\theta$ scheme adds modest demand of $2K\mathcal{M}$ multiplications per stage for $L-1$ stages to reconstruct and cancel MUI estimates. JAPSIC $M_{u}$ requires ranking of channel powers of all APs, adding further $M \log_{2}(M)$ computations to select $M_{u}$ APs. MMSE schemes require inversion of matrices of size $M\times M$ for each user, leading to $\mathcal {O}\big(K^{3}\big)$ multiplications. MMSE-SIC demands $\approx K/2$ times the efforts of MMSE. To give some numbers: for $K = 40$, and $M = 100$, we find
computations required for: MMSE-SIC $=11284000$,
and MMSE $=568000$ operations while for JAPSIC $\theta = 38000$, and JAPSIC
$M_{u} = 38660$ assuming $M_{u} = \mathcal{M} = M/2$, and $L = 10$. The MF \cite{IEEEngo:cfmr}  uses 
$KM=4000$ operations. With UC using $M_{u}=50 $, this reduces to $2660$.

In terms of backhaul signalling, JAPSIC requires $M$ APs to send total of $\tau_{c}M$ complex scalars to the CPU every coherence period that is same as in \cite{IEEEngo:cfmr}. No channel correlation matrices needed to be known at the CPU unlike in \cite{IEEEbjo:cfmmse}, thus saving signalling efforts to send further $KM/2$ complex scalars.
\vspace{-2mm}
\section{Numerical Results}
For further comparisons, we use simulations assuming the following setup. We take an $1$ km × $1$ km area with
$K$ users randomly distributed within the area and $M=100$ single
antenna APs. All users transmit with power $p_{k} = 100$ mW,
carrier center frequency is $2$ GHz and system bandwidth $20$
MHz, thermal noise power is $-174$ dBm/Hz and noise figure at
APs of $5$ dB, $\tau_{c} = 200$, $\rho=4$ and $\tau_{p} = K/\rho$.
\begin{table}[!t]
\caption{Comparison Of Computational Efforts Of Different Cell-Free
Massive MIMO Schemes For Each Detection Cycle}
\centering
\begin{tabular}{c|c}
\hline
{Scheme} & {Initial Filtering, Post Processing of Signal Vectors}\\
\hline
\hline
MMSE \cite{IEEEbjo:cfmmse} & $KM+K((M^{2} + KM) + M)$\\
\hline
MMSE-SIC \cite{IEEEbjo:cfmmse} & $KM+K/2 \{ K((M^{2} + KM) + M)\}$\\
\hline
JAPSIC $\theta$ & $K\mathcal{M}+ (L-1)2K\mathcal{M}$\\
\hline
JAPSIC $M_{u}$ & $KM_{u}+ (L-1)2KM_{u}+M\log_{2}(M)$\\
\hline
UC \cite{IEEEbuzzi:cfuc} & $KM_{u}+M\log_{2}(M)$\\
\hline
MF \cite{IEEEngo:cfmr} & $KM$\\
\hline
\end{tabular}
\vspace{-4mm}
\end{table}
Figure 1 shows the cumulative distribution function (CDF) graphs of the sum SEs achieved by the JAPSIC
schemes against the MF \cite{IEEEngo:cfmr}, UC \cite{IEEEbuzzi:cfuc}, MMSE \cite{IEEEbjo:cfmmse} and MMSE-SIC \cite{IEEEbjo:cfmmse} with $K=40$ and $M= 100$ under the same centralized system setup for fair comparisons. As expected, JAPSIC schemes give much higher sum SE as $\theta/M_{u}$ is lowered/increased. With this change, the JAPSIC algorithms pick more APs with stronger channels
to give better estimates of users’ data and this knowledge is aptly exploited in multiple stages to cancel MUI and refine all users’ data estimates. 
JAPSIC with $\theta = 0.01/M_{u}=10$ far outperform MF/UC,  but are still inferior to MMSE; note however that the modest increase of complexity of JAPSIC may still be justifiable. With $\theta =0.0001/M_{u}=50$, they outperform MMSE as well as MMSE-SIC while performing close to the full interference cancellation (F-IC) using about half the APs. This can be attributed to the ability of JAPSIC to collect most of the useful signals and cancel MUI without enhancing estimation errors and noise (\ref{sinr_japsic_a}).



\begin{figure}[!t]
\centering
\includegraphics[width=3.5in, height=1.4in]{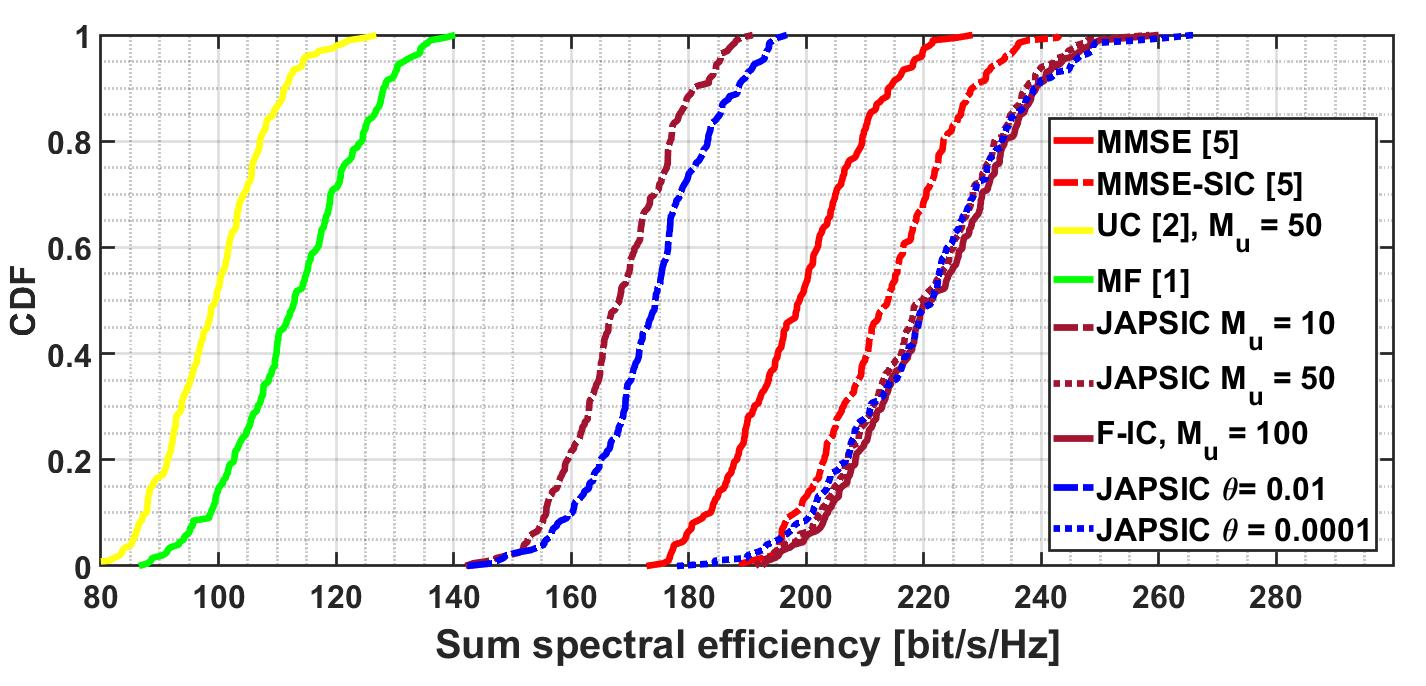}
\caption{Comparison of CDF of sum spectral efficiencies for the proposed JAPSIC schemes against other CF schemes for $K = 40$, $M=100$; where JAPSIC $M_{u}$ with $M_{u}=\{10,50\}$ and JAPSIC $\theta$ with $\theta=\{0.01,0.0001\}$ giving $\mathcal {M}=\{10.55,43.20\}$, respectively, are used.}
\label{fig_sim}
\vspace{-4mm}
\end{figure}

In Figure 2 we show the relative SINRs of MMSE and JAPSIC $M_{u}$ to asssess the impact of possible pilot contamination as given in (\ref{sinr_mmse_a}) and (\ref{sinr_japsic_a}). We use a simplified model assuming all users' channels follow uncorrelated Rayleigh distribution with equal variances and introduce diffferent degrees of channel estimation errors $\sigma^{2}_{c}$. Under a small $\sigma^{2}_{c}=-20 dB$, at low SNR region, MMSE achieves higher SINR compared with JAPSIC with $M_{u}=50$. However under higher SNR, as the channel estimation error enhancement of MMSE (\ref{sinr_mmse_a}) become more visible, JAPSIC with $M_{u}=50$ can outperform it. With $\sigma^{2}_{c}=-10 dB$, the gain of  JAPSIC over MMSE diminishes. 
\begin{figure}[!t]
\centering
\includegraphics[width=3.5in, height=1.4in]{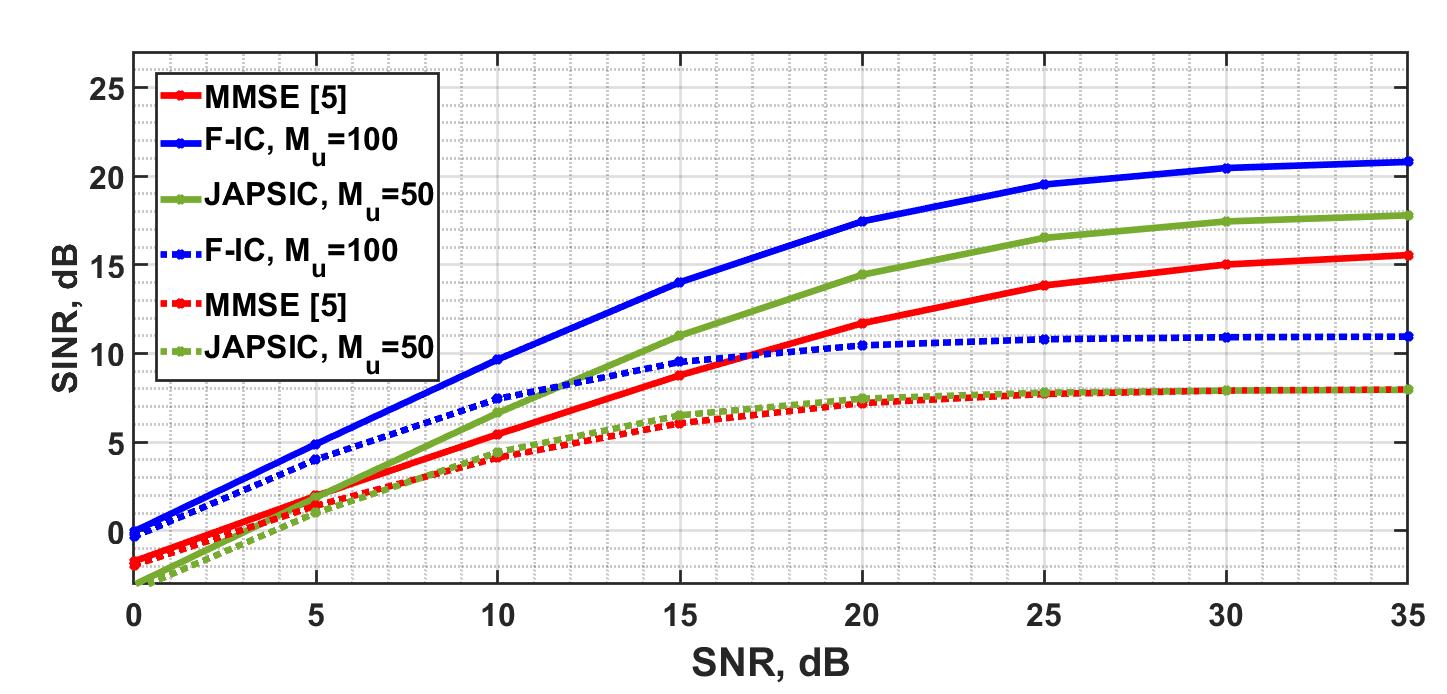}
\caption{Achievable SINR of JAPSIC $M_{u}$ against MMSE under channel estimation errors of $\sigma^{2}_{c}=-20dB$ (solid lines) and $\sigma^{2}_{c}=-10dB$ (dotted lines)
with $K=80$ and $M=100$.}
\label{fig_sim}
\vspace{-4mm}
\end{figure}

\vspace{-3mm}

\section{Conclusions}
We demonstrated a new low complexity and high capacity approach called JAPSIC that employs joint process of selective combining of AP signals and multistage 
interference cancellation as an attractive alternative to MMSE based CF massive MIMO. With
analyses and numerical results, we verified substantial gains both in terms of spectral and computational efficiencies that
merits the scheme proposed. For example, at a computational demand of just $0.3\%$ of the MMSE-SIC, it can achieve higher sum SE
of 220 bits/s/Hz compared with 214 and 199 for the MMSE — and almost double the MF that achieves
only 112 bits/s/Hz. For the future work, it will be interesting to expand and analyse the JAPSIC algorithms under different channel and user loading environments and assess the performance with different optimization methods.

\vspace{-2mm}


%





\ifCLASSOPTIONcaptionsoff
  \newpage
\fi

\end{document}